\documentclass[preprint,12pt]{elsarticle}

\usepackage[utf8]{inputenc}  % For UTF-8 encoding
\usepackage{float}           % Improved control over float placement
\usepackage{epsfig}          % For including EPS images
\usepackage{natbib}          % For bibliography
\usepackage{amsmath}         % Math environments
\usepackage{times}           % Times New Roman font
\usepackage{subfigure}       % For subfigures
\usepackage{graphicx}        % For including images
\usepackage{color}           % Text color
\usepackage{soul}            % Highlighting
\usepackage{dsfont}          % Special math fonts
\usepackage{verbatim}        % Verbatim text
\usepackage{xcolor}          % Extended colors
\usepackage{array, multirow} % Tables with multi-row and multi-column
\usepackage{booktabs}        % Enhanced tables
\usepackage{bm}              % Bold math symbols
\usepackage{hyperref}        % Hyperlinks
\usepackage{algpseudocode}   % Algorithms
\usepackage{algorithm}       % Floating algorithm environment

\hypersetup{
    colorlinks=true,
    linkcolor=blue,
    citecolor=blue,
    urlcolor=blue
}

\DeclareUnicodeCharacter{0301}{\'}  % Fix Unicode accents

\begin{document}

\begin{frontmatter}  % Required for elsarticle template

\title{AI-Driven Control of Chaos: A Transformer-Based Approach for Dynamical Systems}

\author{David Valle}
\ead{david.valle@urjc.es}

\author{Rub\'{e}n Capeans}
\ead{ruben.capeans@urjc.es}

\author{Alexandre Wagemakers\corref{cor1}}
\ead{alexandre.wagemakers@urjc.es}

\author{Miguel A.F. Sanju\'{a}n}
\ead{miguel.sanjuan@urjc.es}

\cortext[cor1]{Corresponding author}

\address{Nonlinear Dynamics, Chaos and Complex Systems Group, Departamento de F\'{i}sica, Universidad Rey Juan Carlos, Tulip\'{a}n s/n, 28933 M\'{o}stoles, Madrid, Spain}

\begin{abstract}
Chaotic behavior in dynamical systems poses a significant challenge in trajectory control, traditionally relying on computationally intensive physical models. We present a machine learning-based algorithm to compute the minimum control bounds required to confine particles within a region indefinitely, using only the first iterations of diverging orbits as required information of the system. This model-free approach achieves high accuracy, with a mean squared error of $2.88 \times 10^{-4}$ and computation times in the range of seconds. The results highlight its efficiency and potential for real-time control of chaotic systems.
\end{abstract}
\end{frontmatter}

\section{Introduction}
The study of chaos and the development of methods to predict and control chaotic systems have a rich history, with significant progress made over the past few decades. Early approaches focused on understanding the mathematical foundations of chaos, leading to the identification of key properties such as Lyapunov exponents and strange attractors, which describe the system's sensitivity to initial conditions and the structure of chaotic trajectories~\cite{peitgen1992chaos}. These foundational concepts paved the way for the development of control strategies aimed at stabilizing chaotic systems~\cite{Ott1990, Shinbrot1990, Braiman1991, Pyragas1992, Shinbrot1993, Grebogi1997}.

Building on these early successes, researchers have developed various methods to control chaotic systems, with significant focus on managing chaotic transients. These transients are periods of chaotic motion that occur temporarily before a system stabilizes or escapes to another attractor. Managing chaotic transients is essential in systems where maintaining a chaotic regime offers significant benefits. For example, in mechanical systems, chaotic vibrations can improve the efficiency of energy harvesters by broadening the range of resonance frequencies and increasing energy extraction~\cite{litak2015regular}. Similarly, in ecological models, chaotic dynamics can help prevent population collapses and maintain ecosystem stability~\cite{capeans2014when}. Among the most effective techniques developed to achieve these goals is partial control~\cite{sabuco2012dynamics,capeans2017partially}, with safety functions and safe sets as their core elements~\cite{capeans2019new}.

The safety function is a powerful tool for controlling systems with chaotic transients governed by difference equations. It quantifies the minimum control required to confine chaotic trajectories within an arbitrary region of phase space for a given number of iterations~\cite{capeans2022controlling,Capeans2024}. While this approach represents a significant advancement in the control of chaotic systems, the computation of safety functions remains highly resource intensive. This challenge becomes particularly evident in systems with a high number of dimensions, as well as in scenarios requiring real-time decision making. As a result, this computational bottleneck limits the broader adoption of safety functions in fields that demand rapid and efficient solutions.

In this work, we aim to address these challenges by incorporating machine learning techniques to estimate safety functions. The primary goal is to reduce the computational burden of safety function calculations while maintaining accuracy, making the approach feasible for real-time and large scale applications. 

Our methodology involves collecting datasets from unstable systems with chaotic transients, training models with varying levels of available information, and evaluating their performance. By focusing on generalization, we create predictive models of safety functions for a wide range of systems without requiring fine tuning. Once the models have been trained, the control of a system only requires samples of diverging trajectories of the uncontrolled system and the actual state. This approach not only improves efficiency but also expands the practical utility of safety functions in controlling chaotic transients.

The structure of this paper is as follows. Section~\ref{Theoretical_background_sec} reviews the current state of the art in the computation of safety functions and discusses the role of machine learning in addressing these challenges. Section~\ref{Methodology_sec} details the data collection, model design, and training processes. In Sec.~\ref{Results_sec}, we evaluate the predictive models, examining their performance and efficiency across different scenarios. Finally, Sec.~\ref{Conclusion_sec} summarizes our findings, discusses their implications, and outlines potential avenues for future work.

\section{Theoretical background}
\label{Theoretical_background_sec}

Transient chaos refers to the temporary chaotic behavior exhibited by a system before it stabilizes or escapes to a specific region. This phenomenon presents unique challenges when combined with the intrinsic instability of chaotic systems and the unavoidable presence of noise, which exists in any real experimental setup or dataset. Maintaining a small control under these conditions may seem unrealistic. However, it was discovered that a region $Q$, which contains a chaotic saddle, also includes a special subset where chaotic trajectories can be sustained with a minimal amount of control~\cite{sabuco2012dynamics,capeans2018partial}. This subset, known as the safe set, depends on both the noise strength affecting the system and the available control.

The partial control method focuses on keeping chaotic trajectories confined within the safe set using minimal control even when noise is present. Surprisingly, partial control can find safe sets where the magnitude of control needed to prevent the chaotic orbit from escaping is smaller than the noise itself. A recent approach~\cite{capeans2019new} enhances the partial control method by introducing a safety function, which automatically computes the minimum control required for each point in the region $Q$ to remain within the desired bounds. This method has been applied to well known chaotic systems such as the Hénon and Lozi maps, demonstrating that chaotic dynamics can be controlled with minimal intervention~\cite{capeans2022controlling}.

Consider the noisy, partially controlled discrete map:
\begin{equation}
q_{n+1}=f(q_n, \xi_n) + u_n ,\qquad |\xi_n| \le \xi,
\label{eq:controlled_map}
\end{equation}
where $f$ is the deterministic map, $\xi_n$ is an additive disturbance bounded by the noise amplitude $\xi$ and $u_n$ is the control effort applied at each iteration.

Without control most trajectories eventually leave the interval $Q$ displaying transient chaos. To quantify how much control is required at every point
$q_i\in Q$, we introduce the sequence of functions $\{U_k\}$ generated by the recursion

\begin{equation}
\begin{aligned}
&U_{0}(q_i)   = 0 , \qquad \forall\, q_i \in Q ,\\[2pt]
& U_{k+1}(q_i) = g(U_k,q_i,\xi).
\end{aligned}
\label{Safety_function}
\end{equation}

Here, the operator $g$ implements the classical safety function algorithm, as detailed in~\ref{Appendix_A}, which is the target of optimization in this paper.  For a finite horizon $k$, the value $U_k(q_i)$ represents the minimum control upper bound required to ensure that the trajectory starting at $q_i$ remains within the set $Q$ for at least $k$ iterations. The corresponding sequence of $k$ controls applied satisfies that independently of the realization of the noise: 
\begin{equation}
\max (|u_1|, |u_2|, \dots, |u_k|) \leq U_k(q_i).
\label{eq:controlleqsafety}
\end{equation}

To ensure that the system’s trajectory remains indefinitely within the set $Q$, we consider the limit of the sequence $\{U_k\}$ as $k \to \infty$.  Fortunately, we do not need to compute infinite functions, as the presence of noise induces convergence of this sequence after a finite number of steps. That is, there exists some finite value of $k$ such that $U_{\infty}=U_{k+1}=U_{k}$. This function $U_{\infty}$ is called \textit{the safety function}. For each $q_i$ the value $U_\infty(q_i)$ gives the smallest control upper bound that guarantees confinement in $Q$ for an arbitrarily long time. 

Once the safety function $U_\infty$ has been determined, the control of a trajectory follows immediately. With the only knowledge of the actual perturbed state $f(q_i, \xi_n)$, we have to evaluate every admissible controlled image:
\[
q_j = f\bigl(q_i, \xi_n\bigr) +\; u_n , \qquad q_j\in Q,
\]
and choose, among all feasible controls $u_n$, the one that minimizes the worst-case upper bound:
\begin{equation}
\label{eq_control}
\underset{q_j \in Q}{\min}\;
\Bigl(
  \max\bigl(|u_n|,\, U_\infty(q_j)\bigr)
\Bigr).
\end{equation}

Applying this rule at every iteration guarantees that the entire trajectory starting at $q_i$ remains in $Q$ while satisfying the bound $\lvert u_n\rvert\le U_\infty(q_i)$.

Moreover, if we focus only on those states that can be stabilized with an amplitude not exceeding a prescribed value $u$, we define \textit{the safe set} $\mathcal{S}(u)$ as:
\begin{equation}
\mathcal{S}(u)=\bigl\{\,q\in Q : U_{\infty}(q)\le u\,\bigr\}.
\end{equation}

Thus, any orbit starting in $\mathcal{S}(u)$ can be maintained within $\mathcal{S}(u)$ indefinitely by applying a control input not exceeding $u$ at each iteration of the map, as defined in Eq.~\eqref{eq:controlled_map}.

\begin{figure}[!h]
\vspace{-55pt}
\hspace{-30pt}
\includegraphics[scale=0.17]{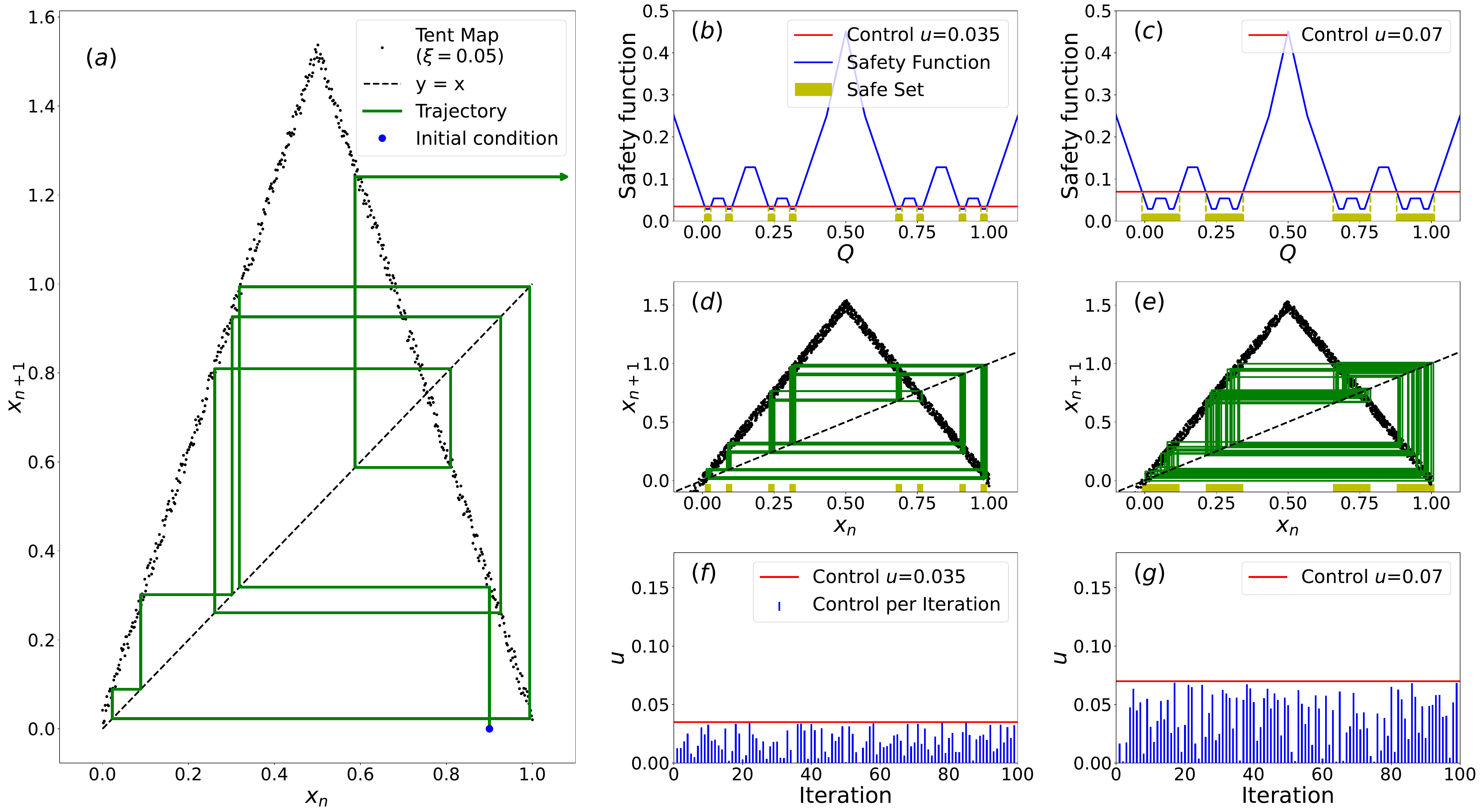}
\caption{\textbf{Illustration of transient chaos and the application of the safety function for the tent map with noise $\bm{\xi = 0.05}$ in the interval $\bm{Q = [0, 1]}$.} Panel $(a)$ shows a non controlled chaotic trajectory of the noisy map, eventually escaping the region $Q$. Panels $(b)$ and $(c)$ display the safety function $U_\infty$ for the map in $(a)$, with maximum control values arbitrarily set to $u = 0.035$ and $u = 0.07$, respectively (red line). The yellow regions represent the safe set, corresponding to areas where the safety function lies below the maximum control value $u$, allowing trajectories to remain indefinitely when control is applied. Panels $(d)$ and $(e)$ show two trajectories confined to the safe set, with control $u_n \leq u$ applied at each iteration, contrasting the escape seen in $(a)$. Panels $(f)$ and $(g)$ illustrate the control values applied during the first $100$ iterations, confirming that they remain consistently below the selected maximum $u$. This demonstrates the effectiveness of the safety function in maintaining trajectories within region $Q$.}
\label{SafetyExample}
\end{figure}

An example of the application of this method is shown in Fig.~\ref{SafetyExample}, where we apply this control method to the slope-3 tent map with a noise $\xi_n \leq \xi$. The map is defined as:

\begin{equation}
x_{n+1} =
\begin{cases} 
3x_n + \xi_n + u_n & \text{for } x_n \leq \frac{1}{2}, \\ 
3(1 - x_n) + \xi_n + u_n & \text{for } x_n > \frac{1}{2}.
\end{cases}
\end{equation}

Where $x_n$ is the state and $u_n \leq u$ is the control applied at each iteration. This map exhibits transient chaos within the interval $Q = [0, 1]$. In Fig.~\ref{SafetyExample}(a), we illustrate how an orbit undergoes chaotic motion until it eventually exits $Q$.

Figures~\ref{SafetyExample}(b-c) depict the same safety function associated with this map using a noise $\xi = 0.05$. To identify a safe set $\mathcal{S}(u)$, we extract the set of initial conditions for which the safety function is below a predefined value $u$ (represented by a red line), we arbitrarily set $u=0.035$ in Fig.~\ref{SafetyExample}(b) and $u= 0.07$ in Fig.~\ref{SafetyExample}(c). The resulting safe set, highlighted in yellow, is where the trajectory is confined indefinitely.

Figures~\ref{SafetyExample}(d-e) demonstrate how the orbit iterates within the map, bouncing within the regions of the safe set indefinitely, in contrast to the case shown in Fig.~\ref{SafetyExample}(a) where no control is applied and the orbit diverges. Finally, Figs.~\ref{SafetyExample}(f-g) show the value of the control applied during the first $n = 100$ iterations.

Computing Eq.~\ref{Safety_function} poses a significant challenge due to its iterative nature. When applied to systems with high-dimensional state spaces or evaluated across numerous initial conditions, the computational cost scales significantly. This complexity can pose challenges for real-time applications or large scale implementations, necessitating efficient computational methods to make the process feasible in practice.

Machine learning offers a viable solution to these challenges by approximating the convergence behavior. This approach has been successfully employed in various fields of physics to optimize complex processes. For instance, neural networks have been used to represent quantum wavefunctions, accelerating the estimation of ground states in many body quantum systems~\cite{carleo2017solving}. Convolutional neural networks have been applied to characterize the complexity of the basins of attraction in chaotic systems, significantly outperforming classical algorithms in processing speed~\cite{Valle2022,Valle2024}. Deep reinforcement learning has also been leveraged to tackle complex control problems in plasma physics, enabling the control of magnetic fields and stabilization of plasma in nuclear fusion reactors~\cite{degrave2022magnetic}.

These applications show how machine learning can reduce the computational demands of iterative estimation processes and optimize control tasks in high-dimensional physical systems, highlighting the potential of similar methods for convergence estimation in dynamical systems like the one under study. A major advancement in the field has been the introduction of transformer models, as described by Vaswani et al.~\cite{vaswani2017attention}. Transformers revolutionized natural language processing by replacing recurrent and convolutional layers with a self-attention mechanism, capturing dependencies in sequences regardless of distance. This innovation improved parallelizability and reduced training times compared to earlier architectures like Long Short-Term Memory networks~\cite{hochreiter1997long} and Convolutional Neural Networks~\cite{lecun1998gradient}.

The strengths of transformer models extend beyond natural language processing to domains with complex relationships and demanding computations. These capabilities make transformers particularly suited to improve the estimation of convergence and control. By learning intricate patterns in data, transformers offer a promising approach to approximating safety functions, and at the same time to address the computational requirement of real-time control. While this study focuses on one-dimensional systems, the adaptability of transformers suggests they could be extended to higher dimensional problems in future work. The subsequent sections detail the application of transformers to our problem.

\section{Methodology}
\label{Methodology_sec}
This work has been fully implemented in Python, with all scripts and data publicly available in a \href{https://github.com/RedLynx96/Estimation_of_safety_functions}{GitHub repository} \cite{RedLynx96_Estimation_of_safety_functions}.

To generate the datasets, we developed a Python script that creates one-dimensional pseudorandom physical models. For these systems, we collect samples of orbits that iterate within a predefined region $Q$ until they escape, along with the reference safety function for this region. Additional details about the generation of pseudorandom functions, sampled orbits, and safety functions can be found in \ref{Appendix_B}.

\begin{figure}[!h]
\vspace{-80pt}
\hspace{-20pt}
\includegraphics[scale=1.8]{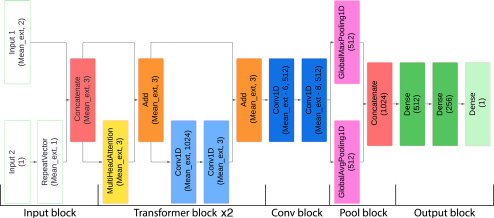}
\caption{\textbf{Architecture of the neural network used in this research.} The model includes two input branches: Input 1, representing the time series of sampled orbits with array length \texttt{Mean\_ext}, and Input 2, the initial condition for which the control value is calculated. Input 2 is repeated and merged with Input 1 before being processed through a stack of two transformer blocks, a convolutional block, a pooling block, and a final output block. Each layer's output shape is shown below its name. Key components include Multi Head Attention layers to capture dependencies in the time series, Conv1D layers for local pattern extraction, and pooling layers to reduce dimensionality. Dense layers with Dropout finalize the prediction of the minimum control value required to keep the initial condition within the safe set indefinitely.}
\label{architecture_fig}
\end{figure}

The transformer-based neural network architecture used in this work is illustrated in Fig.~\ref{architecture_fig}. The model takes time series data as input, representing multiple concatenated orbits of a single dynamical system, along with an initial condition from region $Q$, and outputs the estimated safety function value corresponding to that initial condition within the dynamical system.

The architecture comprises a stack of two transformer blocks, followed by a convolution block, a pooling block, and a final output block of fully connected layers. This configuration was chosen as it empirically provided the best performance during preliminary experiments. The two transformer blocks effectively capture complex, long range dependencies in the input time series, while the subsequent convolution block complements this by extracting localized patterns. This balance allows the model to generalize well and accurately predict the safety function.  

To evaluate the impact of input length on model accuracy, we trained separate models with the same architecture but different input vector lengths. Specifically, models were trained on time series data with input sizes of $50$, $100$, $250$, $500$, $1000$, and $2000$ sampled points.

\clearpage

During each training epoch, $51200$ data samples were dynamically generated instead of using a fixed dataset. These samples came from various dynamical systems, with diversity introduced by randomizing the map's governing function, orbit sampling, and noise values in each iteration. The training lasted for $500$ epochs, with a batch size of $64$, processing a total of approximately $26$ million samples. The system used for training included two Nvidia Quadro RTX $5000$ GPUs running in parallel and a Ryzen Threadripper Pro $3945$WX CPU.

All network weights and biases were randomly initialized and optimized using TensorFlow's~\textit{Adam} optimizer~\cite{kingma2014adam}, configured with the following parameters: learning rate $= lr_0 = 1.5 \cdot 10^{-4}$, $\beta_1 = 0.9$, $\beta_2 = 0.999$, $\epsilon = 10^{-8}$, $\text{decay} = 0$, and $amsgrad = \text{true}$. A warmup phase was introduced at the start of training to adapt the learning process to the transformer architecture. The tuning of the learning rate is described below:
\begin{equation}
lr_t =
\begin{cases} 
lr_0 \cdot \frac{t}{W_e} & \text{for } t \leq W_e \\
lr_0 \cdot d^{\frac{t - W_e}{T - W_e}} & \text{for } t > W_e.
\end{cases}
\end{equation}

Here $lr_t$ denotes the learning rate at epoch $t\geq 1$, starting at an initial value of $lr_0$. The learning rate increases linearly during the $W_e = 6$ warmup epochs, followed by a decay phase controlled by a decay factor $d = 0.2$. This warmup phase is an effective step for the training of transformer models, as it helps stabilize optimization by gradually increasing the learning rate during the initial training phase, preventing issues with large gradients~\cite{vaswani2017attention}.

The model was optimized to minimize a custom Weighted Mean Square Error (Wmse) loss function, defined as:
\begin{equation}
\text{Wmse} = \frac{1}{M} \sum_{i=1}^{M} w_i \cdot (y_{t,i} - y_{p,i})^2,
\end{equation}
where $M = 51200$ represents the total number of observations per epoch, $y_{t,i}$ is the target value, and $y_{p,i}$ is the predicted value for observation $i$. A custom weight $w_i$ is applied to encourage the neural network to overestimate predictions, defined as:
\begin{equation}
w_i = 
\begin{cases} 
10 & \text{if } (y_{t,i} - y_{p,i}) > 0 \\
1 & \text{if } (y_{t,i} - y_{p,i}) \leq 0.
\end{cases}
\end{equation}

This weighting approach introduces a bias for the overestimation of the safety function. Underestimation can lead to an orbit escaping early the region $Q$. Since the safety function defines the minimum control bound needed to confine the orbit, underestimations pose a critical risk. To mitigate this, the training process imposes a significantly higher penalty on underestimations, prompting the model to prioritize larger predictions. This conservative choice for the estimation lead to a satisfactory trade-off in performance as we will show in the next section.

\section{Results}
\label{Results_sec}
As detailed in Sec~\ref{Methodology_sec}, the transformer-based model illustrated in Fig.~\ref{architecture_fig} was trained on different amounts of sampled data. We trained $6$ models, each with progressively more data than the last. This approach simulates real world scenarios where data availability may be limited.

\begin{figure}[!h]
\vspace{-80pt}
\hspace{-40pt}
\includegraphics[scale=0.40]{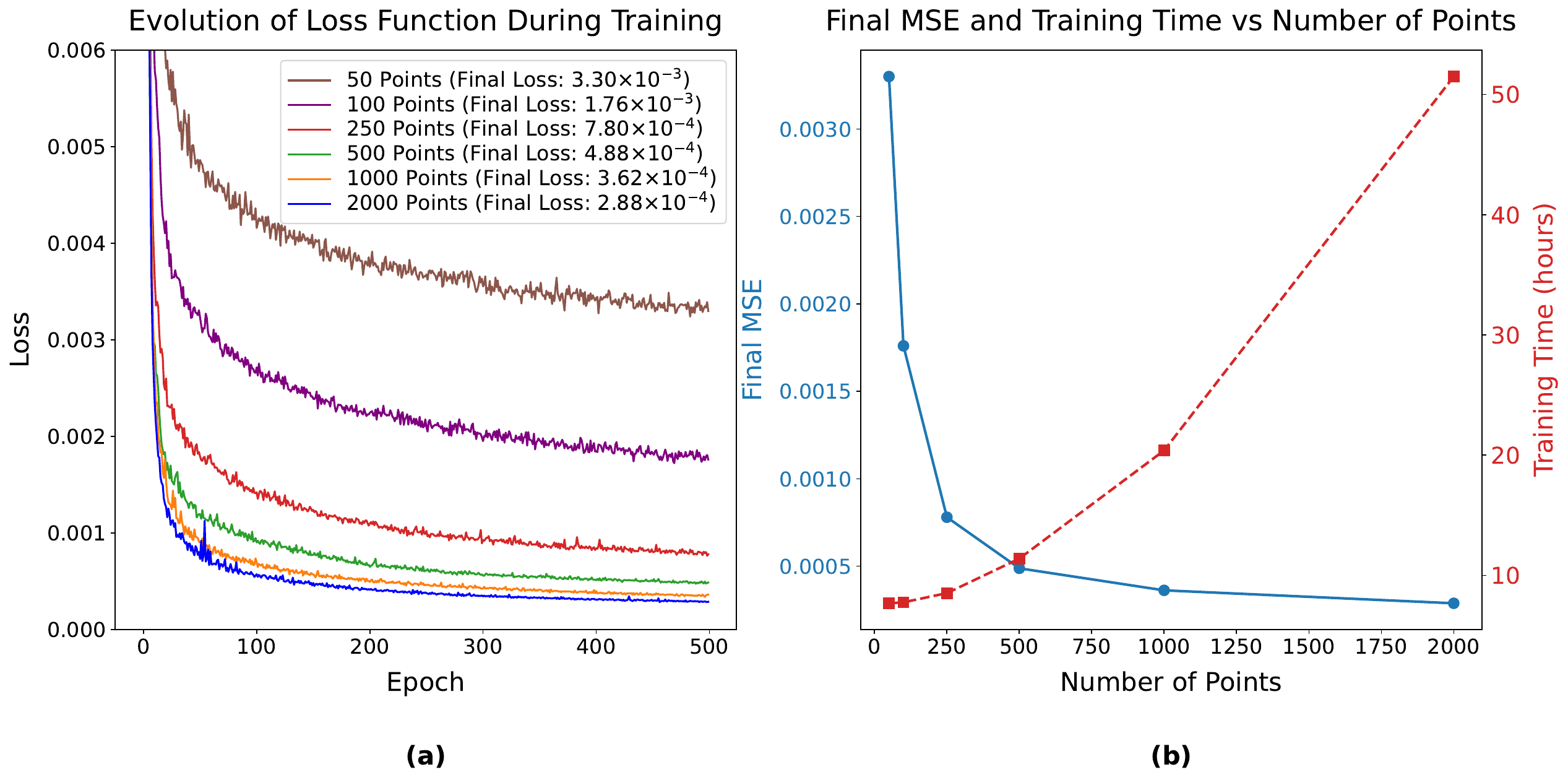}
\caption{\textbf{Comparison of training results for different sample sizes in the transformer-based model.} (a) Illustrates the evolution of the loss function during training for each model, demonstrating faster convergence and lower loss as the model uses more sampled points. (b) emphasizes the trade-off between model accuracy and computational cost. With an increase in the amount of information, the final Mean Squared Error (MSE) decreases significantly, but the training time increases sharply, particularly for larger sample sizes.}
\label{training_results}
\end{figure}

Figure~\ref{training_results} presents the training results, where the prediction error decreases as the model receives more data and undergoes further training. However, this also significantly increases training time, primarily due to the transformer blocks acting as a bottleneck. Training these blocks requires approximately $O(N^2 \cdot d)$ time, where $N$ represents the number of sampled points, and $d$ is the data dimensionality. Compared to the computation times of other blocks, the transformer blocks impose the primary constraint on training speed.\newpage

\begin{figure}[H]
\vspace{-110pt}
\hspace{-65pt}
\includegraphics[scale=0.48]{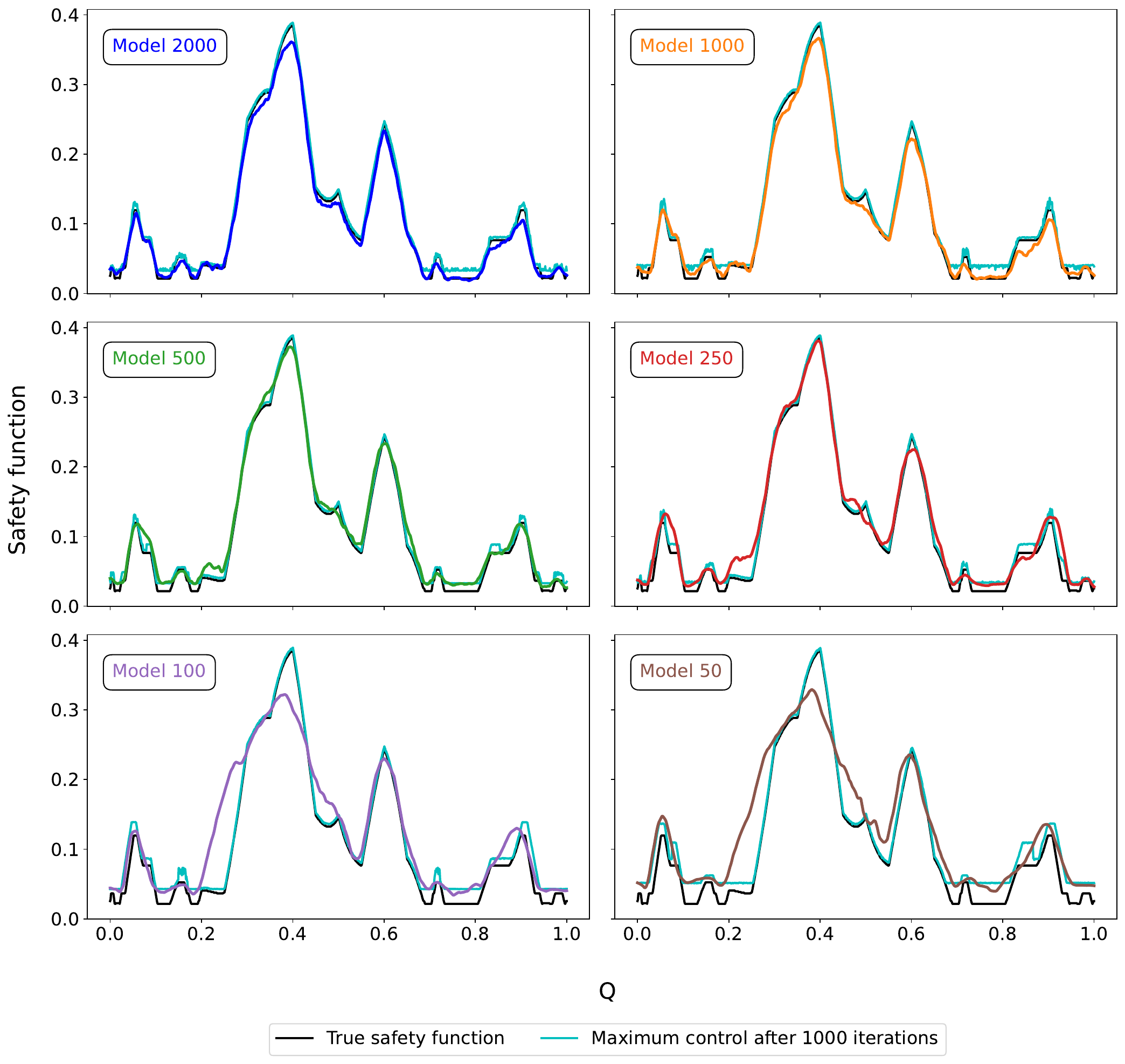}
\caption{\textbf{Comparison of predicted safety functions with the true safety function across various models.} Each subplot illustrates the similarity of the predictions with their ground truths. As the number of sampled points increases, the predicted safety function aligns more closely with the true safety function. The thin blue line denotes the maximum control for each initial condition over $1000$ iterations, calculated by applying Eq.~\ref{eq_control} and using the predicted safety function as $U_\infty$.}
\label{SafetyFunctionComparison} 
\end{figure}

Figure~\ref{SafetyFunctionComparison} illustrates the neural network’s accuracy by comparing the true safety function of a map generated randomly with the safety functions predicted by the trained models. The neural network predicts the value of the safety function $U_\infty(x_k)$ for a single input state $x_k$. To reconstruct the complete safety function, $1000$ initial conditions were sampled from the region $Q$, and the same time series data was used as input for each prediction. Networks trained on a larger number of points produce predictions that closely approximate the true safety function, though they tend to slightly underestimate the curve. Conversely, models trained on fewer points are more likely to overestimate the safety function.

In addition to the predicted safety functions, each plot includes a cyan curve representing the maximum control across all initial conditions in $Q$ over $1000$ iterations following the controls computed with Eq.~\ref{eq_control} at each step and using as $U_\infty$ the predicted safety function. Notably, the maximum of the control time series typically aligns closely with the true safety function. However, in regions of minimum control, a lower bound emerges where all orbits exhibit the same maximum control value.

\begin{figure}[H]
\vspace{-110pt}
\centering
\includegraphics[scale=0.62]{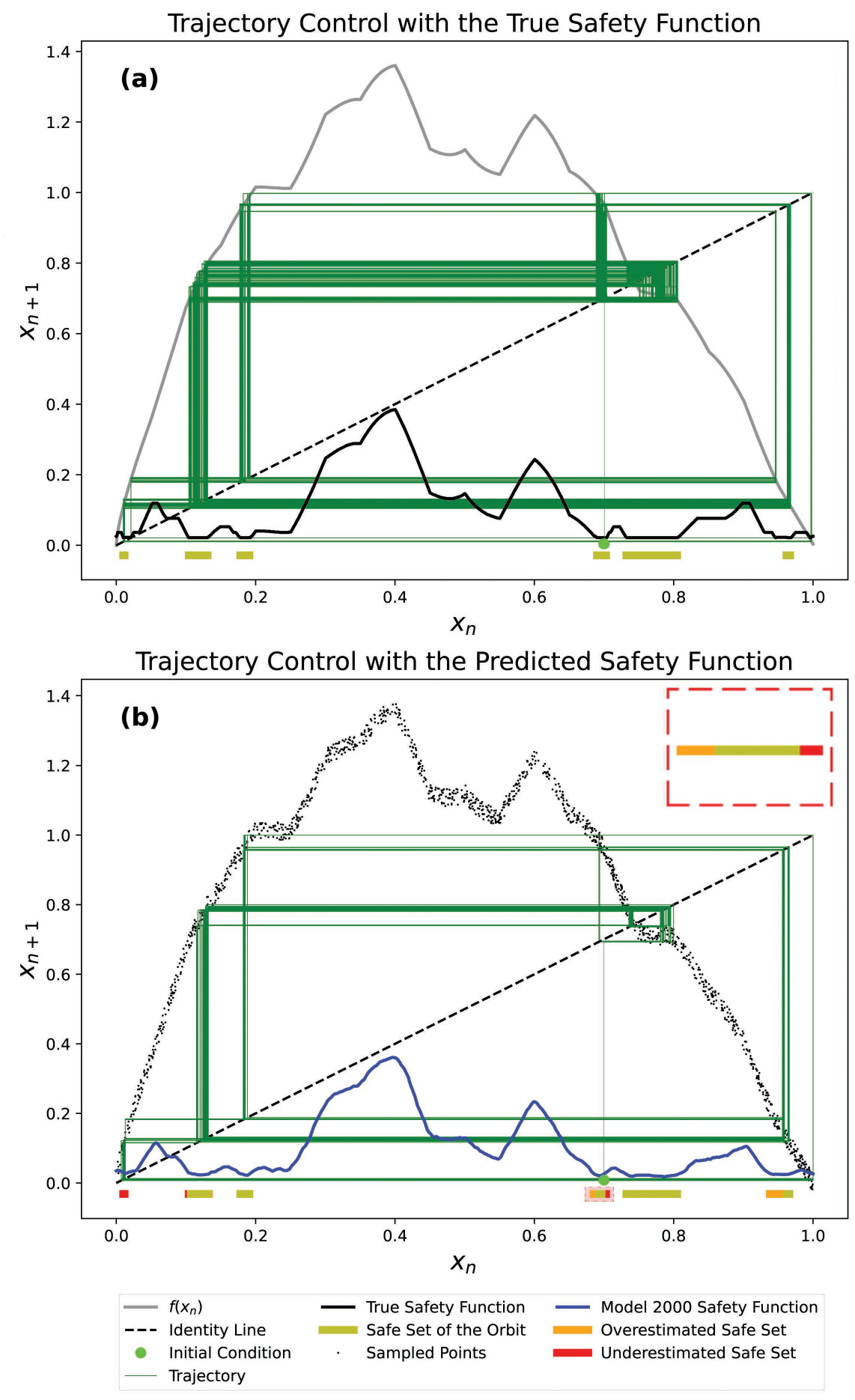} 
\caption{\textbf{Comparison of the trajectory following the control values of the true safety function (a), and the predicted safety function from the model trained with $\mathbf{2000}$ points (b)}. The plots depict an orbit initialized at $x_n = 0.7$, iterated $1000$ times, showing the trajectory, safety function, and derived safe set. In panel (a), the safe set, derived from the true safety function, ensures indefinite control of the orbit. In panel (b), errors in the predicted safety function lead to overestimated regions (orange) and missing regions (red) in the safe set.} 
\label{SafetyOrbitVsPredictedOrbit} 
\end{figure}

Figure~\ref{SafetyOrbitVsPredictedOrbit} represents the observed behavior in an orbit with initial condition $x_n = 0.7$ iterated $1000$ times with the control computed with Eq.~\ref{eq_control} at each step. The difference between the two time series is due to the method for estimation of the safety function. In the horizontal axis, the intervals of the safe set appear in pale yellow. The safe set includes all regions in $Q$ where the safety function is less or equal to this value.

Using the true safety function in Fig.~\ref{SafetyOrbitVsPredictedOrbit}(a), the safe set identifies regions where the orbit remains stable indefinitely under the suggested control. In Fig.~\ref{SafetyOrbitVsPredictedOrbit}(b), the predicted safety function, derived from a model trained with $2000$ points, introduces misalignments in the safe set. Overestimation of the safety function leads to underestimation of the safe set (red regions), while underestimation of the safety function causes overestimation of the safe set (orange regions). A zoomed in view near $x_n = 0.7$, shown in the top right corner of Fig.~\ref{SafetyOrbitVsPredictedOrbit}(b), highlights these discrepancies.

Underestimated regions (red) correspond to missing parts of the safe set where the control is overestimated. This means the orbit could remain stable, but the predicted safe set excludes these areas. For example, in the interval $[0.003, 0.017]$, the predicted control is higher than necessary, and this part of the true safe set is omitted. However, this does not cause significant issues because applying the overestimated control still keeps the orbit stable. Since all orbits eventually pass through this interval, the overestimated control becomes dominant for the entire safe set.

Overestimated regions (orange) occur where the orbit is predicted to bounce but does not in reality. This happens because the actual control required is higher than the predicted value. If the initial condition is chosen in these regions, the orbit may escape the predicted safe set. Alternatively, if a higher control than predicted were applied, the orbit could eventually bounce into parts of the true safe set that the method fails to capture. Furthermore, the absence of this interval in the predicted safe set creates a feedback loop where the orbit's trajectory appears less stable than it truly is, introducing inaccuracies in both the predicted trajectory and the required control. This highlights the critical sensitivity of the safe set to inaccuracies in the safety function, emphasizing the need for improved prediction models to minimize such discrepancies.

A significant challenge in this model-free approach lies in accurately capturing the lower control regions of the true safety function, particularly when data is sparse. In such scenarios, the estimated safety function often diverges from the true safety function in these regions. This discrepancy arises because sharp variations in the lower control zones depend on data points that are frequently absent during sampling, hindering precise detection and representation. This phenomenon aligns with previous studies on dynamical systems subject to minimal disturbance, which demonstrate that as control decreases, the safe set of orbits within a region $Q$ forms a Cantor-like structure, eventually vanishing from the system~\cite{capeans2019new}.
\newpage

\begin{figure}[!h]
\hspace*{+10pt}
\includegraphics[scale=0.6]{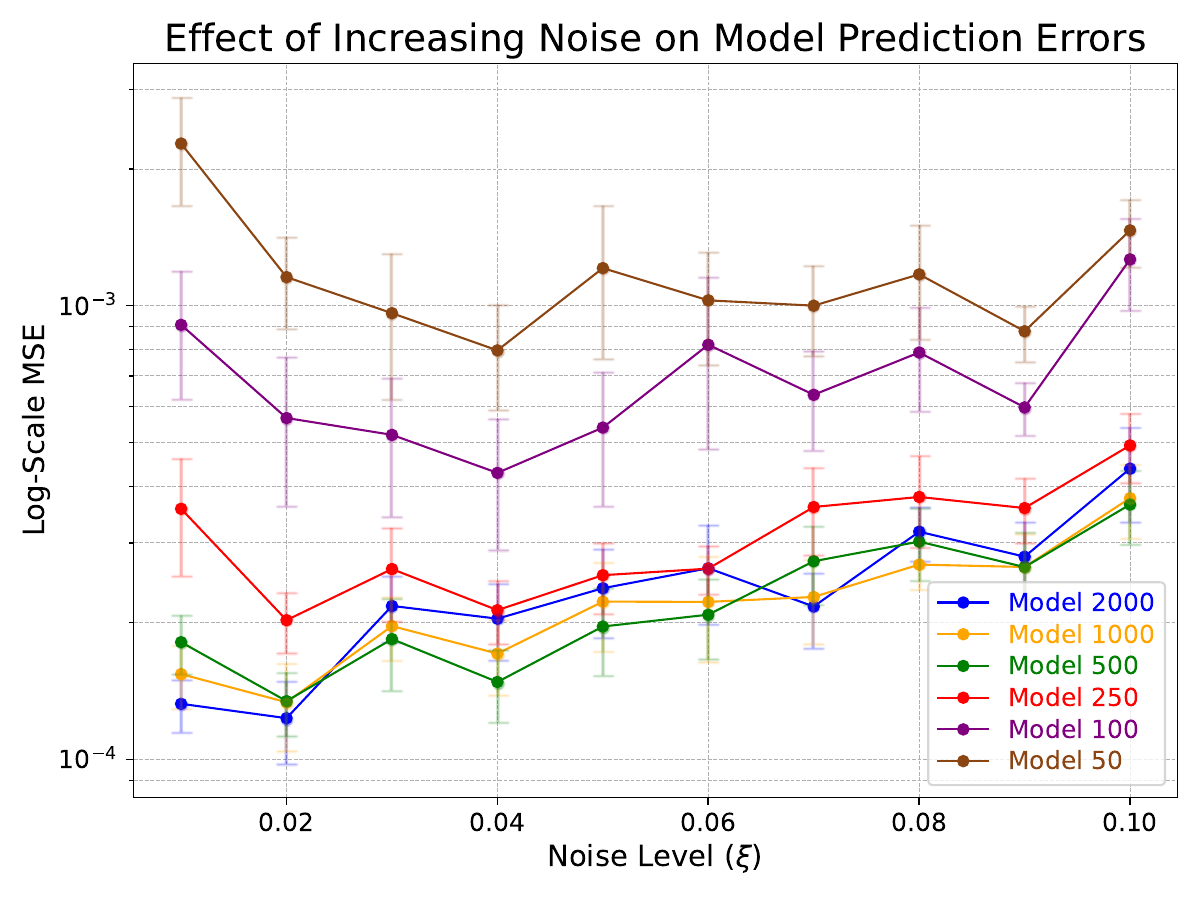} 
\caption{\textbf{Mean Squared Error (MSE) as a function of increasing noise values ($\bm{\xi}$) for each model.} The plot shows that the MSE increases as the noise level rises, reflecting reduced predictive accuracy under noisier maps. Models with higher information from the map, such as Model 2000 and Model 1000, generally exhibit lower MSE values and greater resilience to noise compared to smaller models like Model 50 and Model 100, which show both higher MSE and greater variability. Error bars denote the standard deviation of MSE, indicating performance fluctuations for each model size at different noise levels.} 
\label{mse_noise}
\end{figure}

Regarding the performance of the trained models, Fig.~\ref{mse_noise} illustrates the evolution of the mean squared error (MSE) in estimating safety functions as the noise value $\xi$ in the source maps increases. This analysis involved the computation of $1500$ safety functions derived from random maps. Specifically, for each of the $100$ distinct values of $\xi$ ranging from $0.001$ to $0.1$, $15$ safety functions were generated using maps that shared the same value of $\xi$. For each safety function, the mean MSE was calculated in two steps: first, determining the average MSE across all $1000$ points defining the predicted safety function; and second, averaging these MSE values over the $15$ safety functions.

The results demonstrate that the models estimate safety functions more accurately for maps with lower noise upper bounds, with performance deteriorating as the noise value increases. Nevertheless, the predictions remain reliable across the entire range of noise levels. This trend aligns with a training induced bias, as described in \ref{Safety_func_computation_appendix}, where systems with lower noise were prioritized to enhance the estimation of the lower control regions of the safety function.

To conclude the analysis of the aforementioned dataset, Table~\ref{tab:model_performance_summary} provides a detailed summary of each training model's performance, including the mean computations time per safety function with $1000$ points and its standard deviation, as well as the mean MSE of the dataset and its standard deviation. For comparison, we also show the computation time of the safety functions obtained through the classical method. As explained in \ref{Appendix_A}, these functions are computed using Eq.~\ref{True_Noisy_Safety_function} and evaluated on the same set of points $Q$ as the predicted safety functions. The average computation time is calculated over the $1500$ safety functions of the last dataset and over $50$ iterations of the equation, assuming convergence is reached.

\begin{table}[h!]
\centering
\scalebox{1.0}{
\begin{tabular}{|c|c|c|}
\hline
\textbf{Model} & \textbf{Mean MSE} & \textbf{Average Prediction Time (s)} \\ \hline
$2000$ & $(2.14 \pm 1.36)\cdot 10^{-4}$ & $3.84 \pm 0.65$ \\ \hline
$1000$ & $(2.22 \pm 1.30)\cdot 10^{-4}$ & $1.35 \pm 0.29$ \\ \hline
$500$ & $(2.27 \pm 1.44)\cdot 10^{-4}$ & $0.68 \pm 0.18$ \\ \hline
$250$ & $(3.26 \pm 2.34)\cdot 10^{-4}$ & $0.63 \pm 0.22$ \\ \hline
$100$ & $(7.05 \pm 5.91)\cdot 10^{-4}$ & $0.62 \pm 0.13$ \\ \hline
$50$ & $(1.32 \pm 1.08)\cdot 10^{-3}$ & $0.62 \pm 0.18$ \\ \hline
\hline
$\text{Classical Algorithm}$ & & $0.75 \pm 0.06$ \\ \hline
\end{tabular}}
\caption{\textbf{Model performance summary:} Increasing the number of sampled points generally improves accuracy (lower MSE) but extends computation time. Models with $1000$ or $500$ points achieve similar MSE to the $2000$ point model under evenly distributed noise. Additionally, models with fewer than $500$ points show minimal computation time variation, likely due to hardware constraints discussed in Section~\ref{Methodology_sec}. They also outperform the classical safety function estimation algorithm in terms of computation time. This table, in conjunction with Fig.~\ref{mse_noise}, highlights the trade-off between model accuracy, prediction time, and the effects of model size on prediction error across different noise levels.}
\label{tab:model_performance_summary}
\end{table}

As expected, increasing the amount of information sampled from a system, results in longer prediction times and reduced MSE. However, for models processing $500$ points or fewer, no significant variation in computation time was observed. This consistency in computation time is likely attributed to the hardware specifications used for executing the models, as discussed in Sec.~\ref{Methodology_sec}. This lower threshold reflects the computational overhead of the programs that can be optimized to meet a application requirement.

Finally, we explored the impact of varying the amount of sample data on the prediction of safety functions. Specifically, we evaluated whether it is more effective to discard data and use a model trained with fewer points or to apply padding and use a model trained on a larger dataset.

\begin{figure}[H] 
%\vspace{-70pt}
\centering
\includegraphics[scale=0.69]{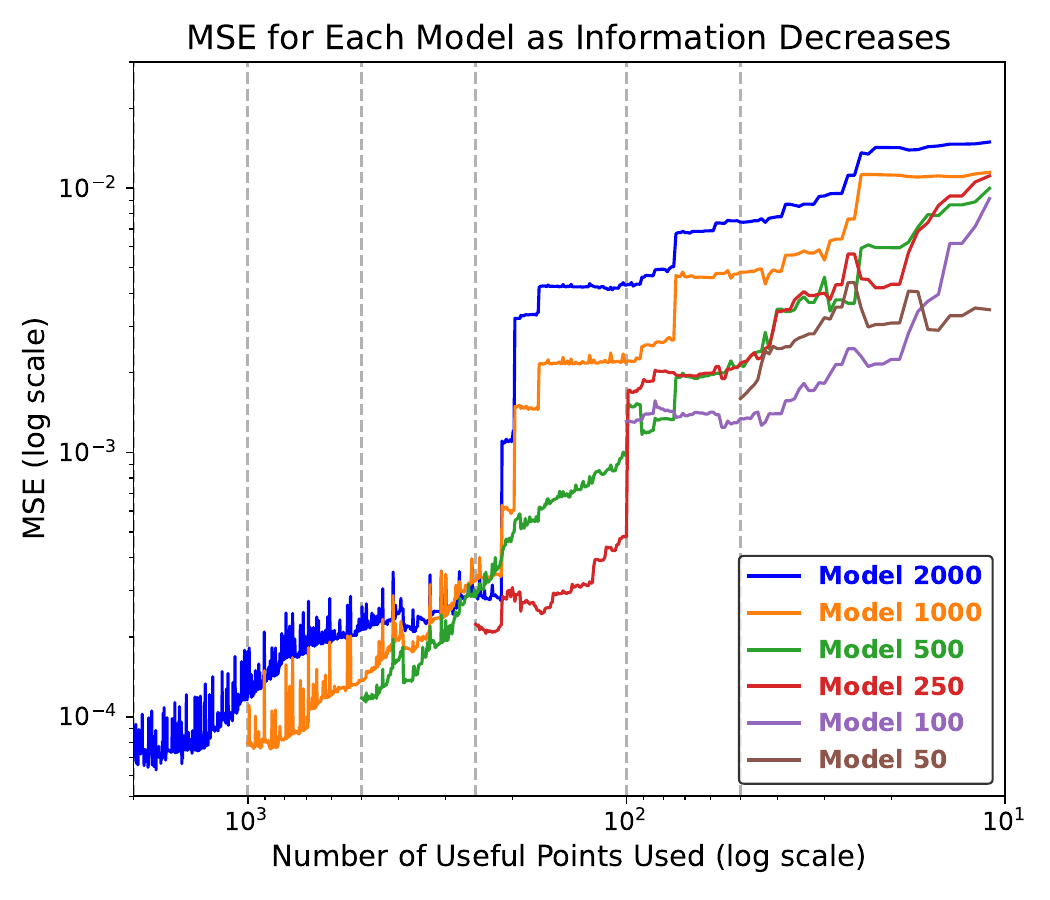} 
\caption{\textbf{Evolution of the Mean Squared Error (MSE) as the number of useful points for prediction decreases.} The MSE increases with increasing padding, with significant performance shifts occurring at specific sample sizes. Models trained on smaller datasets perform better when the sample size aligns with or is smaller than their training range, as they are better adapted to limited information. Notably, the model trained on $100$ points performs robustly across a wide range of small sample sizes and becomes the most effective for fewer than $50$ points. This demonstrates the ability of smaller models to adapt to sparse data scenarios, where larger models struggle due to insufficient training exposure for such conditions.}
\label{MSEvsNPoints}
\end{figure}

Figure~\ref{MSEvsNPoints} illustrates the evolution of the MSE for each model as padding increases. The time series data analyzed are the same as those used to compute the safety functions shown in Fig.~\ref{SafetyFunctionComparison}. As expected, the MSE increases as the number of useful sampled points decreases, highlighting the challenges posed by reduced information.

Key transitions occur in this example at $200$ points for models trained on $2000$ and $1000$ points, and at $100$ points for the model trained with $250$ points. These jumps occur because the model’s training was heavily biased toward larger datasets, causing abrupt performance degradation when applied to significantly smaller samples, where the training data distribution does not align with the test conditions.

\section{Conclusion}  
\label{Conclusion_sec}  

In this study, we address the fundamental control objective of transforming transient chaos into permanent chaotic motion using only time series. To meet this goal, we propose a transformer-based framework designed to predict the safety functions of dynamical systems without the need for an explicit physical model. This model-free approach offers notable advantages over traditional algorithms. Specifically, our framework achieves competitive accuracy while relying only on the time series of the data. Whereas classical methods require prior knowledge of the system map and set an upper bound of disturbance, our approach, once trained, merely needs a representative time series taken from the unstable dynamical system, to approximate the safety function. Consequently, the control of an orbit depends only on the current perturbed state and the predicted safety function, enabling real-time control.

The quantitative results emphasize the efficiency and reliability of this approach. When trained on $2000$ sample points, the model achieved a mean squared error (MSE) of $2.88 \times 10^{-4}$, closely approximating the values produced by classical methods. Furthermore, this level of accuracy remained stable across a range of noise levels, as demonstrated in Fig.~\ref{mse_noise}, highlighting the robustness of the model under varying input conditions.

Interestingly, the analysis also revealed differences in noise sensitivity based on the amount of sampled points. Models trained with larger information, such as Model $2000$ and Model $1000$, displayed greater sensitivity to noise, with MSE increasing as noise levels rose. In contrast, smaller models showed MSE fluctuations around their mean values, without a clear trend of worsening under noisy conditions.

This suggests that larger sampling models, while effective at identifying lower control regions within safety functions, may exhibit reduced robustness to noise. In contrast, smaller sampling models focus on capturing the overall structure of the safety function and tend to deliver more consistent performance under noisy input conditions.

A key strength of the transformer-based approach is its significantly lower computational complexity compared to classical methods. The classical computation of safety functions, as defined in \ref{Appendix_A}, involves an iterative process with a computational complexity of $\mathcal{O}(K \cdot M \cdot N \cdot 2^d)$, where $K$ is the number of iterations required for the safety function to converge, $M$ is all the possible disturbance scenarios affecting a point, $N$ is the number of neighbors per point, and $d$ is the dimensionality of the system. This iterative nature makes classical methods computationally prohibitive for higher dimensional systems, especially as the number of neighbors or required iterations increases.

Our proposed model integrates transformer blocks, convolutional layers, pooling layers, and dense layers to achieve efficient computation with reduced complexity. The dominant computational cost arises from the $\mathcal{O}(N^2 \cdot d)$ term of transformer blocks, supplemented by the $\mathcal{O}(N \cdot d \cdot k)$ term from convolutional layers (where $k$ is the kernel size), while pooling and dense layers contribute minimally. This architecture substantially reduces computation time compared to classical methods, particularly in cases with large $N$ or high $d$. This efficiency makes the transformer-based framework suitable for real-time control applications.

Despite its strengths, the model has certain limitations. Its prediction range is confined to $Q = [0,1]$, with an image upper bound of $1.5$ and a maximum noise threshold of $0.1$. Extending its applicability requires data normalization to match input conditions with the training domain, adding preprocessing steps and complexity. Moreover, while the model excels in regions requiring substantial control, it tends to overestimate control requirements in low control zones. This bias arises from the underrepresentation of rare, low noise orbits in the training dataset. Addressing this issue involves augmenting the dataset with samples that better capture low control dynamics.

We also explored the trade-off between data volume, predictive accuracy, and computation time for safety functions composing of $1000$ points. As shown in Table~\ref{tab:model_performance_summary}, increasing sample size generally improves accuracy and robustness but at the cost of increased training time and computational demands. These results suggest that selecting an optimal balance between sampled points and computational cost can yield efficient and accurate performance tailored to specific applications. Further research could explore strategies such as adaptive sampling and padding to enhance model performance with limited or imbalanced data.

In conclusion, the proposed transformer-based framework offers a robust, model free solution for estimating the minimum control value required to maintain an initial condition of a map exhibit transient chaos within a region $Q$ indefinitely.

By achieving sub $2.88 \times 10^{-4}$ MSE accuracy and drastically reducing prediction times compared to classical methods, the approach demonstrates strong potential for real-time control applications in low and high-dimensional systems. Future research should focus on improving the model's scalability, addressing current limitations, and exploring new strategies for dimensional scalability. These developments could further extend the utility of this approach across higher dimensions and real data experiments.  

\appendix
\section{Classical computation of safety functions}
\label{Appendix_A}
This appendix provides a detailed methodology for the classical computation of safety functions. These functions play a critical role in  extending the partial control method, enabling the precise determination of minimal control bounds necessary to sustain trajectories within a \mbox {predefined region $Q$ of the phase space.}

The safety function, $U(q)$, represents the minimum control magnitude required to prevent a trajectory, starting at point $q \in Q$, from escaping $Q$ under bounded disturbances. This function generalizes the concept of the safe set, offering a more direct computation of the minimal control strategy.

\subsection{Computation with disturbances}
For any dynamical system with noise, the evolution of a point $q$ is governed by the equation
\begin{equation} 
q_{n+1} = f(q_n, \xi_n) + u_n, \quad |\xi_n| \leq \xi, |u_n| \leq u, 
\end{equation} 
where $f(q)$ defines the system dynamics, $\xi_n$ represents the disturbance affecting the system at each iteration, bounded by $\xi$, and $u_n$ represents the applied control at each iteration, constrained by a maximum allowable magnitude $u$. The disturbances $\xi_n$, while bounded, can create multiple possible future states for each current state $q_n$. Therefore, the computation of the safety function must account for all potential disturbance scenarios, making the process inherently more complex than in the deterministic case.

The algorithm begins with an initialization phase. The region $Q$ is discretized into a grid of $N$ points, and the initial safety function $U_0[q_i]$ is set to zero for all grid points $q_i$, reflecting that no control is applied initially. This provides the foundation for iteratively building up the control function, step by step.

For each grid point $q_i$, the map $f(q_i, \xi_s)$ is evaluated across all possible disturbance scenarios $\xi_s$, where $s$ indexes a set of $M$ representative disturbance scenarios affecting $q_i$. This approach explicitly accounts for all potential impacts of the disturbances, resulting in a collection of disturbed images for each $q_i$ that represents the range of possible next states under the influence of noise.

\begin{figure}[H]
\vspace{-80pt}
\hspace{-15pt}
\includegraphics[scale=0.6]{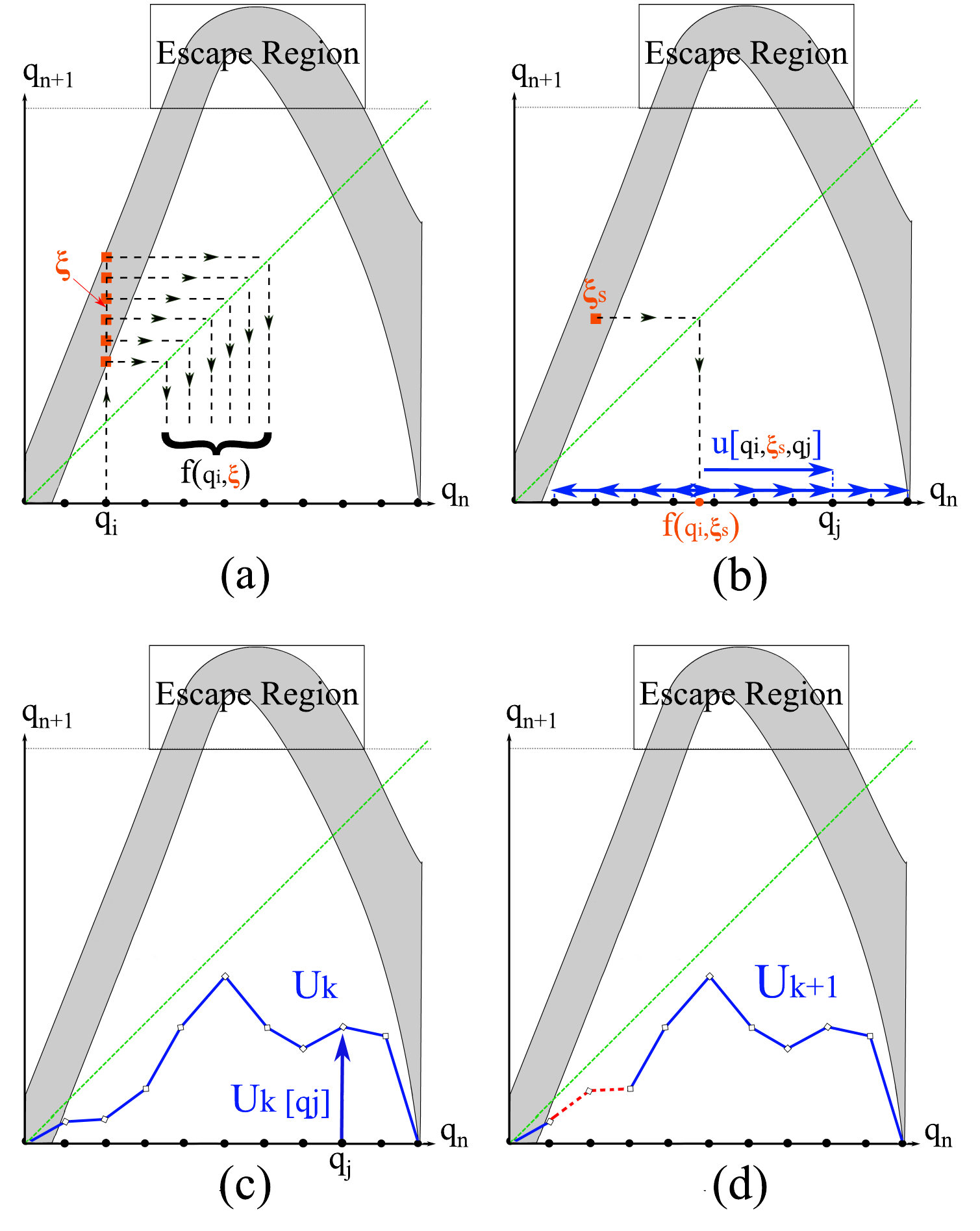}
\caption{\textbf{Step-by-step computation of the safety function \bm{$U_{k+1}[q_i]$} under disturbances.} This diagram illustrates how the safety function is updated at a grid point $q_i$. (a) A set of disturbances $\xi_s$ is applied to $q_i$, producing the disturbed images $f(q_i, \xi_s)$. (b) From each disturbed point, we compute the control required to reach every other grid point $q_j \in Q$. (c) Each control value is compared with the current safety value $U_k[q_j]$, and for each disturbance, the target $q_j$ requiring the least combined effort is selected. (d) Among all disturbances, the worst-case (maximum) of these minimal efforts is used to update the safety function at $q_i$ as $U_{k+1}[q_i] = \max \big(u[q_i, \xi_s, q_j], U_k[q_j] \big)$.}
\label{Algoritmo_explicativo}
\end{figure}

Recursive computation is then performed to update the safety function iteratively. For each iteration $k+1$, and for each grid point $q_i$, the algorithm evaluates all possible transitions $q_i \to q_j$ under all disturbance scenarios $\xi_s$. The computation involves calculating the control $u[q_i, \xi_s, q_j]$ required to move from the disturbed image $f(q_i, \xi_s)$ to $q_j$. For each combination of $\xi_s$ and $q_j$, the algorithm computes the upper bound of the control required, $\max(u[q_i, \xi_s, q_j], U_k[q_j])$, where $U_k[q_j]$ represents the control needed to sustain the trajectory starting at $q_j$ for $k$ further iterations. This process ensures that the control strategy considers not only the immediate step but also the long term trajectory within $Q$. Figure~\ref{Algoritmo_explicativo} depicts this iterative control process 

The recursive update for $U_{k+1}[q_i]$ then selects the worst case disturbance impact by taking the maximum over all disturbance scenarios $\xi_s$ and simultaneously minimizes the overall control bound by selecting the optimal next state $q_j$. This step is expressed mathematically as: 
\begin{equation} 
U_{k+1}[q_i] = \max_{1 \leq s \leq M} \bigg( \min_{1 \leq j \leq N} \big( \max_j \big(u[q_i, \xi_s, q_j], U_k[q_j]) \big) \bigg). 
\label{True_Noisy_Safety_function} 
\end{equation} 

The iterative process continues until $U_k[q_i]$ converges for all $q_i$, indicating that further iterations produce no significant changes. At this point, the algorithm outputs the safety function $U_\infty[q_i]$, which represents the minimal control required for each grid point $q_i$ to remain in $Q$ indefinitely, under all possible disturbance scenarios.

The following pseudocode serves as a guide to compute the true safety function of any dynamical system.

\begin{algorithm}[H]
	\begin{algorithmic}[]
		\Statex
		Given the $N$ points $q_i \in Q$ and their corresponding $M_i$ disturbed images $f\big(q_i,\xi_s\big)\;$, the safety function $U_\infty$ can be computed in 			the following way:\\
		\State- Initially set $\;U_0\,[q_j]=0, \; \forall j=1:N, \; \; \; \; k=0.$
		\State
		\hspace{4cm}\While {$U_{k+1}\neq U_k$}
		\State
		\For{$i=1$ to  $N\;$}
		\For{$s=1$ to  $M\;$}
		\For{$j=1$ to  $N\;$}
		\State $\;u\,[q_i,\xi_s,q_j]= \Big|\,f\big(q_i,\xi_s\big)- \;q_j\,\Big|$
		\State
		\State $\;u^*\,[q_i,\xi_s,q_j]=\max\limits_{j}\,\big(\,u[q_i,\xi_s,q_j],\, U_k[q_j]\,\big)$
		\EndFor
		\State $\;u^{**}\,[q_i,\xi_s]=\min\limits_{j}\,\big(\,u^*\,[q_i,\xi_s,q_j]\,\big)$
		\EndFor
		\State $U_{k+1}[q_i]=\max\limits_{s} \big(\,u^{**}\,[q_i,\xi_s]\,\big)$
		\EndFor
		\State  $k=k+1$\normalsize
		\State
		\EndWhile
		\Statex
		\scriptsize{Comments:
			-Note that the values $u[q_i,\xi_s,q_j]$ remain constant throughout each iteration of the while loop; therefore, they should be computed once and stored for reuse.}
		\Statex
	\end{algorithmic}
\end{algorithm}

\subsection{Calculation of safety functions}
\label{Safety_func_computation_appendix}
In our research the estimation of the safety function begins by introducing bounded noise into a random dynamical system. For each point in the region $Q = [0,1]$, the algorithm considers a range of noise values with an upper bound $\xi \in [0.001, 0.1]$. The noise upper bound for each function is sampled from an exponential distribution with rate parameter $\lambda = 0.5$, then scaled to fit within the specified range. This approach biases $\xi$ toward smaller values, allowing the neural network to adapt more effectively to regions with lower control requirements and increased complexity in the safety function.

Once $\xi$ is selected for a given function, random noise $\xi_n \sim \text{Uniform}(0, \xi)$ is systematically added to the function's output, creating a noisy image. This noisy image represents all potential system outputs under the influence of noise, forming the basis for calculating the required controls.

\section{Computation of random functions and safety maps}
\label{Appendix_B}

In this appendix, we describe the methodology used to generate the dataset for training and testing the transformer-based model. The process involves the generation of random one-dimensional dynamical systems, the application of noise to simulate real world disturbances, and the calculation of safety functions to estimate the minimum control required to maintain trajectories within a predefined safe region. Each step is described in detail below.

\subsection{Calculation of random dynamical systems }

The random dynamical systems are constructed by combining base functions, assigning coefficients, and integrating them into a single expression. The base functions are divided into four categories: Set A, Set B, Set C, and Set D. The specific functions in each group are listed in Table~\ref{tab:Base_functions}. These base functions were chosen to capture a wide range of behaviors, including smooth and non smooth dynamics, oscillatory patterns, and discontinuities.

The number of functions to be used $M$ is randomly determined from predefined ranges for each set: $4$ to $8$ from Set A, $2$ to $4$ from Set B, $5$ to $8$ from Set C, and $1$ to $2$ from Set D. These ranges are designed to ensure diversity in the generated functions. Once the random functions $g_i(x)$ are selected from the respective sets, a weight $c_i$ is assigned to each function. These weights are random coefficients uniformly sampled from the range $[0, 1]$. The weighted selected functions are then combined into a single expression $f(x)$ as shown in Eq.~\ref{BaseFuncsGeneralForm}. 
\begin{equation}
f(x) = \sum_{i=1}^M c_i g_i(x).
\label{BaseFuncsGeneralForm}
\end{equation}

Finally, the resulting function $f(x)$ is scaled to ensure its maximum value lies within the interval $[1.2, 1.5]$. This guarantees the existence of an escape region for the set $Q = [0,1]$. This scaling process produces a diverse range of functions that balance complexity and smoothness. Once the random dynamical system is defined, the function's image is calculated for $1000$ values evenly distributed within region $Q$.

\subsection{Time series simulation}

Regarding the system orbits, we iterated $1000$ random initial conditions under the system's dynamics until they either exited region $Q$ or completed $50$ iterations. The time series evolved according to $x_{n+1} = f(x_n) + \xi_n$. If an orbit exits region $Q$, it is truncated and padded with $-1$ values to achieve a consistent length of $50$ points. This process generates a time series with a shape $(1000,50)$ for each random function.

The $(1000, 50)$ shaped arrays were first expanded to $(1000, 50, 2)$ by adding a second column where the original values were shifted by one position, creating sequential pairs of points for each orbit. Next, all $[-1, -1]$ values were filtered out, and the arrays were restructured to represent the initial condition and its iterations until orbit completion, with $[-1, -1]$ marking the end of each orbit. This process resulted in arrays shaped $(\text{Mean\_ext}, 2)$, where the length (Mean\_ext) varied for each sampled map.

After restructuring, the orbits were consolidated, maintaining the $[-1, -1]$ markers to denote divergence or completion. Across $100$ datasets of $51200$ arrays each, the resulting datasets had an average length of $4238.79$ points, a median of $3680.50$ points, a standard deviation of $2016.86$ points, a minimum length of $2408$ points, and a maximum length of $26877$ points. Since the transformer model requires fixed size datasets, each array needed to be either truncated or padded with $[-1, -1]$ values to achieve uniform lengths. For our research, The model's performance in predicting safety functions was evaluated using arrays of lengths $2000$, $1000$, $500$, $250$, $100$, and $50$ points. 

\begin{table}[H]
\vspace{-80pt}
\centering
\resizebox{1\textwidth}{!}{
\renewcommand{\arraystretch}{1.2}
\begin{tabular}{|p{0.45\textwidth}|p{0.45\textwidth}|}
\hline
\textbf{Set A} & \textbf{Set C} \\ \hline
\parbox[t]{\linewidth}{
$x^2 (1 - x)^2$ \\
$\sin^2(\pi x) x (1 - x)$ \\
$(4 + r) x (1 - x)^2$ \\
$\sqrt{x} (1 - x)^2$ \\
$x^3 (1 - x)^3$ \\
$(1 - \cos(\pi x)) x (1 - x)$ \\
$\tanh^2(4x - 2) x (1 - x)$ \\
$(1 - e^{-4x})^2 x (1 - x)$ \\
$(2 + r) \min\left(\frac{x}{c}, \frac{1-x}{1-c}\right)^2 x (1 - x)$
}
&
\parbox[t]{\linewidth}{
$\lvert (1 - e^{-5x}) e^{-5(1-x)} x (1 - x) \rvert$ \\
$\lvert \log(1 + 9x) (1 - x)^2 \rvert$ \\
$\lvert (x - x^2)^2 \sin(2\pi x) \rvert$ \\
$\lvert x^2 (1 - x) \cos^2(\pi x) \rvert$ \\
$\lvert x (1 - x^4) \rvert$ \\
$\lvert e^{-2(x-0.5)^2} x (1 - x) \rvert$ \\
$\lvert (\sin(2\pi x) + 1) x (1 - x) \rvert$ \\
$\lvert (x \log(x + 10^{-5})) (1 - x) \rvert$ \\
$\lvert \sqrt{x} (1 - x^3) \rvert$ \\
$\lvert x^3 (1 - x^5) \rvert$
}
\\ \hline
\textbf{Set B} & \textbf{Set D} \\ \hline
\parbox[t]{\linewidth}{
$\lvert (1 - x^2) \sin^2(\pi x) \rvert$ \\
$\lvert (x - x^2) \cos^2(\pi x) \rvert$ \\
$\lvert x (1 - x) (1 + x - x^2) \rvert$ \\
$\lvert (1 - x) \sin(\pi x) x \rvert$ \\
$\lvert (x (1 - x))^2 \rvert$ \\
$\lvert x^3 (1 - x) (2 - x) \rvert$ \\
$\lvert x (1 - x) e^{-x} \rvert$ \\
$\lvert (1 - x) (1 - \cos(2\pi x)) \rvert$ \\
$\lvert \sin^2(\pi x) x^2 (1 - x) \rvert$ \\
$\lvert (1 - e^{-x}) (1 - x) x \rvert$ \\
$\lvert \cos(\pi x) x^2 (1 - x) \rvert$ \\
$\lvert (x^2 - x) \cos(\pi x) \rvert$ \\
$\lvert (x (1 - x)) \sin^2(4\pi x) \rvert$ \\
$\lvert (x (1 - x)) \cos^2(6\pi x) \rvert$ \\
$\lvert (x (1 - x)) (\sin^2(2\pi x) + \cos^2(2\pi x)) \rvert$ \\
$\lvert (x (1 - x)) e^{-10(x-0.5)^2} \rvert$ \\
$\lvert (x (1 - x)) (1 + \cos(10\pi x)) \rvert$ \\
$\lvert (x (1 - x)) \lvert \sin(8\pi x) \rvert \rvert$ \\
$\lvert (x (1 - x)) (1 - \cos^2(12\pi x)) \rvert$ \\
$\lvert (x (1 - x)) (1 + \sin^2(6\pi x)) \rvert$ \\
$\lvert (x (1 - x)) (1 + e^{-5(x - 0.3)^2} + e^{-5(x - 0.7)^2}) \rvert$ \\
$\lvert (x (1 - x)) (1 + \sin(4\pi x) \cos(4\pi x)) \rvert$
}
&
\parbox[t]{\linewidth}{
$\lvert \sin(2\pi x) x (1 - x) \rvert$ \\
$\lvert x (1 - x) (\text{mod}(10x, 1) > 0.5) \rvert$ \\
$(3 + r) \lvert \text{mod}(10x, 1) - 0.5 \rvert x (1 - x)$ \\
$\lvert \lfloor 10x - 5 \rfloor (1 - x) x \rvert$ \\
$(4 + r) \lvert \cos(\pi x) \rvert x (1 - x)$ \\
$\lvert \lceil 10x - 5 \rceil (1 - x) x \rvert$ \\
$\lvert (\text{mod}(20x, 1) > 0.5) (1 - x) x \rvert$ \\
$\lvert \tan(\pi(x - 0.5)) (1 - x) x \rvert$ \\
$\lvert \text{mod}(\lfloor 5x \rfloor, 2) (1 - x) x \rvert$ \\
$\lvert \text{sawtooth}(2\pi x) (1 - x) x \rvert$ \\
$\lvert \text{sign}(\cos(2\pi x)) (1 - x) x \rvert$ \\
$\lvert \text{mod}(50x, 1) (1 - x) x \rvert$ \\
$\lvert \cos(4\pi x) (1 - x) x \rvert$
}
\\ \hline
\end{tabular}
}
\caption{\textbf{Table of functions for Sets A, B, C, and D}. Set A functions are inspired by well-known chaotic maps and include logistic, sinusoidal, and polynomial components. Set B functions are smooth and represent oscillatory and exponential behaviors. Set C functions add complexity via logarithmic and Gaussian components. Set D functions introduce sharp transitions and discontinuities. Variables $r$ and $c$ are uniformly distributed in $[0,1]$. The functions include standard math operations: $\log$ is base-10 log, $\text{sawtooth}$ is a saw wave, $\text{sign}$ returns the sign of its argument, $\lfloor\cdot\rfloor$ and $\lceil\cdot\rceil$ are floor and ceiling, and $\text{mod}(a, b)$ is the remainder of $a$ divided by $b$.}
\label{tab:Base_functions}
\end{table}

\section*{Acknowledgments}
This work has been financially supported by the Spanish State Research Agency (AEI) and the European Regional Development Fund (ERDF, EU) under Project No.~PID2019-105554GB-I00 (MCIN/AEI/10.13039/501100011033) and PID2023-148160NB-I00  (MCIN/AEI/10.13039/501100011033). 

\bibliographystyle{unsrt}

%\bibliography{Bibliografia}

\end{document}